\documentclass[conference]{IEEEtran}
\IEEEoverridecommandlockouts
\usepackage{cite}
\usepackage{amsmath,amssymb,amsfonts}
\usepackage{algorithmic}
\usepackage{graphicx}
\usepackage{textcomp}
\usepackage{xcolor}
\usepackage{booktabs}
\usepackage{siunitx}
\usepackage{subcaption}

\usepackage[numbers,sort]{natbib}
\usepackage{cite}

\def\BibTeX{{\rm B\kern-.05em{\sc i\kern-.025em b}\kern-.08em
    T\kern-.1667em\lower.7ex\hbox{E}\kern-.125emX}}
\begin{document}

\title{Exploring Traffic Simulation and Cybersecurity Strategies Using Large Language Models
}

\author{\IEEEauthorblockN{1\textsuperscript{st} Lu Gao* \thanks{*Sarayu Varma Gottimukkala, a computer science graduate student at the University of Houston, contributed equally to this work as a co-author.}}
\IEEEauthorblockA{\textit{Department of Construction Management} \\
\textit{University of Houston}\\
Houston, USA \\
lgao5@central.uh.edu}
\and
\IEEEauthorblockN{2\textsuperscript{nd} Yongxin Liu}
\IEEEauthorblockA{\textit{Mathematics Department} \\
\textit{Embry-Riddle Aeronautical University}\\
Daytona Beach, USA \\
liuy11@erau.edu}
\and
\IEEEauthorblockN{3\textsuperscript{rd} Hongyun Chen}
\IEEEauthorblockA{\textit{Civil Engineering Department} \\
\textit{Embry-Riddle Aeronautical University}\\
Daytona Beach, USA \\
Hongyun.Chen@erau.edu}
\and
\IEEEauthorblockN{4\textsuperscript{th} Dahai Liu}
\IEEEauthorblockA{\textit{Department of Graduate Studies} \\
\textit{Embry-Riddle Aeronautical University}\\
Daytona Beach, USA \\
liu89b@erau.edu}
\and
\IEEEauthorblockN{5\textsuperscript{th} Yunpeng Zhang}
\IEEEauthorblockA{\textit{Department of Information Science Technology} \\
\textit{University of Houston}\\
Houston, USA \\
yzhang119@uh.edu}
\and
\IEEEauthorblockN{6\textsuperscript{th} Jingran Sun}
\IEEEauthorblockA{\textit{Center for Transportation Research} \\
\textit{The University of Texas at Austin}\\
Austin, USA \\
jingransun@utexas.edu}
}

\maketitle

\begin{abstract}

Intelligent Transportation Systems (ITS) are increasingly vulnerable to sophisticated cyberattacks due to their complex, interconnected nature. Ensuring the cybersecurity of these systems is paramount to maintaining road safety and minimizing traffic disruptions. This study presents a novel multi-agent framework leveraging Large Language Models (LLMs) to enhance traffic simulation and cybersecurity testing. The framework automates the creation of traffic scenarios, the design of cyberattack strategies, and the development of defense mechanisms. A case study demonstrates the framework's ability to simulate a cyberattack targeting connected vehicle broadcasts, evaluate its impact, and implement a defense mechanism that significantly mitigates traffic delays. Results show a 10.2\% increase in travel time during an attack, which is reduced by 3.3\% with the defense strategy. This research highlights the potential of LLM-driven multi-agent systems in advancing transportation cybersecurity and offers a scalable approach for future research in traffic simulation and cyber defense.

\end{abstract}


\section{Introduction}

\subsection{Importance of cybersecurity in ITS}

Cybersecurity is of critical importance in ITS due to the increasing reliance on technology and the interconnected nature of these systems \citep{agarwal2024enhancing}. As ITS integrates more advanced technologies such as sensors, communication networks, and data analytics to monitor and manage traffic flow, it also becomes more susceptible to cyber threats \citep{chowdhury2023information}. The integrity and reliability of ITS are paramount to ensuring road safety and preventing disruptions, which can lead to significant economic losses.

The growing connectivity of vehicles and infrastructure introduces new vulnerabilities that can be exploited by malicious actors \citep{feng2022cybersecurity}. Cyberattacks can compromise the safety of road users by manipulating traffic signals, injecting false information, or disrupting critical systems. For instance, attackers can manipulate traffic signals, causing accidents, long delays, and severe congestion \citep{perrine2019implications}. Studies have demonstrated that disabling traffic signals can lead to substantial increases in travel delays, with one study reporting a 4.3 times increase in travel time when 26 signals were disabled \citep{dasgupta2022innovative}. Furthermore, attackers may also penetrate the system to falsify traffic data, which can lead to serious consequences such as crashes and severe traffic congestion.

ITS is vulnerable to a range of cyber threats due to the interconnected components such as control systems and data processing platforms \citep{silwal2024assessing}. These vulnerabilities can arise from inadequate security measures, poor network design, outdated software, and weak authentication. Attackers can target various components of the ITS, including signal controllers, vehicle detectors, on-board units (OBUs), and road-side units (RSUs) \citep{arabi2021reinforcement}. Also, the use of wireless communication channels, particularly with connected vehicles, introduces further vulnerabilities \citep{lebaku2025cybersecurity}. Cyberattacks can take various forms, including unauthorized access, denial-of-service attacks, data breaches, and tampering with traffic signals or sensors . Some specific attacks include Sybil attacks, where fake vehicle identities are injected into the network to manipulate traffic data, and data spoofing attacks, where intentionally modified vehicle data is injected into the system \citep{zhao2022evaluating}.


\subsection{Challenges in current cybersecurity testing methods for ITS}

Current cybersecurity testing methods for ITS face several challenges, particularly in the context of dynamic and complex real-world scenarios. Traditional cybersecurity methods often fall short in detecting advanced threats, especially in decentralized networks like V2X (vehicle-to-everything) communication \citep{sun2024genai}. One significant issue is the difficulty in generating diverse and realistic attack scenarios that can effectively evaluate the robustness of these systems. Moreover, the increasing sophistication of cyberattacks, such as false data injection, replay attacks, and stealthy attacks, requires more advanced detection mechanisms. The dynamic nature of traffic environments and the unpredictable interactions of vehicles make it difficult to model and test for all possible attack vectors. Moreover, the large volume of data generated by vehicles adds complexity to the analysis required for real-time threat detection, demanding advanced algorithms and computing resources \citep{hamhoum2024mistralbsm}.

Furthermore, many existing simulation and testing methods lack the flexibility to interact with or manipulate the environment effectively, preventing thorough training and testing of customized scenarios \citep{gao2024laser}. Online, interactive testing of Autonomous Driving Systems (ADS) is also challenging because actors need to react to each other's behaviors, which is difficult to achieve with data collected from real traffic. Furthermore, real-world testing is costly and constrained in scope, making it difficult to cover all potential corner cases \citep{lu2024realistic}. Traditional methods may also lack the ability to account for rare events, such as the presence of emergency vehicles or degraded communication \citep{pang2024illm}. These limitations highlight the need for more advanced, AI-driven solutions that can generate realistic adversarial attacks and enhance the training of detection models.

\subsection{The need for autonomous, multi-agent cybersecurity testing}

Given the challenges discussed above, a potential solution to address these issues is to use multi-agents LLMs to automate cybersecurity testing. Several studies and frameworks have emerged to address the need for automated penetration testing and security simulations \citep{wang2024sands}. These approaches use LLMs and multi-agent systems to simulate the collaborative workflow of human testing teams \citep{kong2025vulnbot}. The frameworks often use a modular design, with specialized agents for different tasks or phases of testing \citep{marantos2024leveraging}. For example, \citet{bianou2024pentest} developed PENTESTAI which includes agents for scanning and searching, exploit validation, and reporting. Similarly, \citet{kong2025vulnbot} developed VulnBot which utilizes agents for reconnaissance, scanning, and exploitation, with a summarizer module to facilitate inter-agent communication. 



\subsection{Current LLM applications in traffic simulation}

Recently, LLMs are being integrated with traffic simulation to enhance various aspects of traffic simulation, analysis, and control \citep{chen2024multimodal}. LLMs can interpret natural language inputs to generate traffic scenarios, road networks, and simulation parameters, which reduces the need for manual coding \citep{li2024chatsumo}. LLMs can also be used to generate green wave control policies for urban arterial roads, which can reduce congestion and improve traffic flow \citep{tang2024large}. They can also help in refining decisions made by RL agents by incorporating real-time information, such as the presence of emergency vehicles \citep{pang2024illm}. 



\section{Methodology}
\begin{figure}[htbp]
    \centering
    \includegraphics[width=1\linewidth]{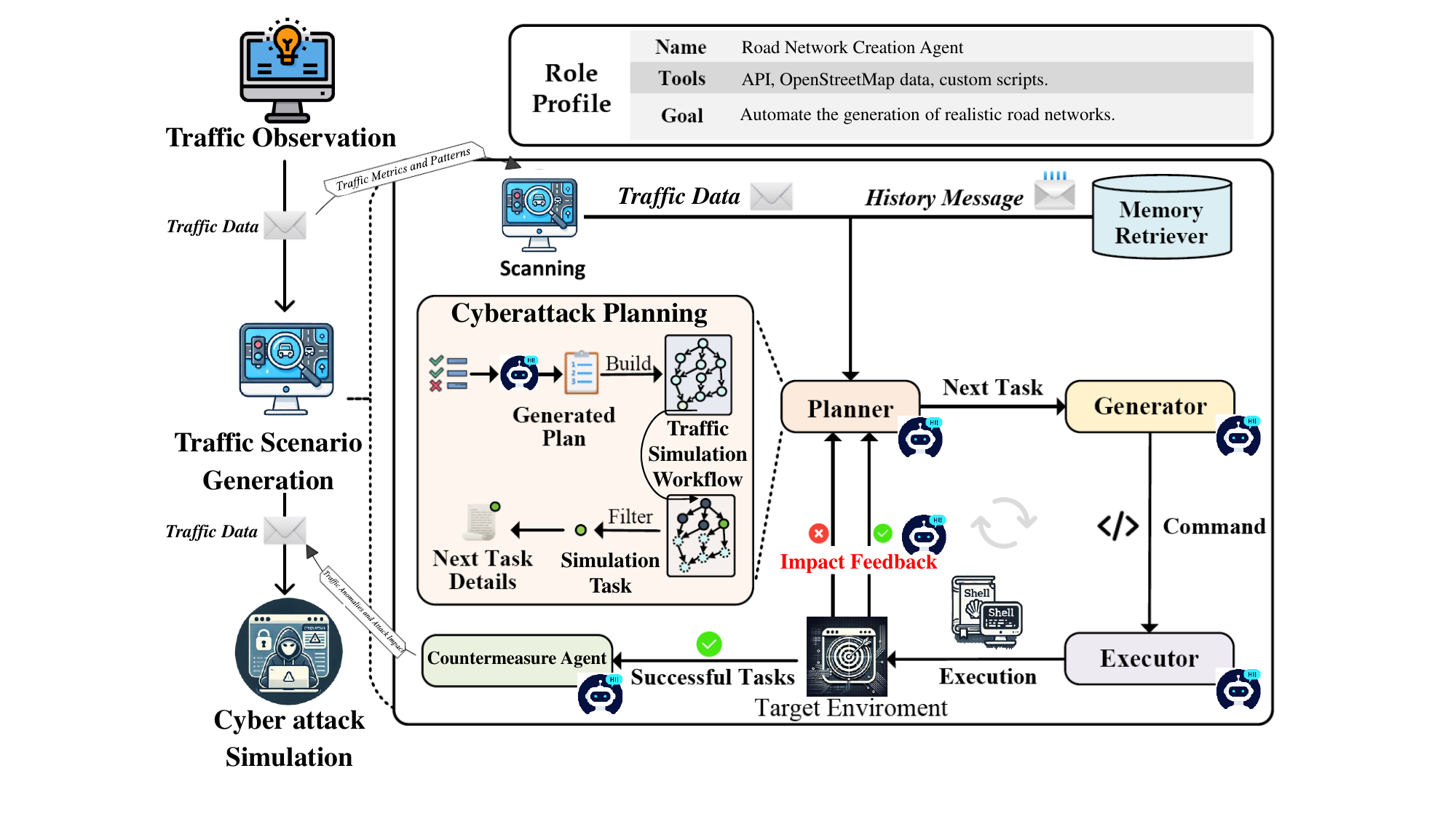}
    \caption{Overview of the Multi-Agents LLM Framework}
    \label{fig:overview}
\end{figure}

In this research, we propose a multi‐agent system architecture (see Figure \ref{fig:overview}) that leverages LLM‐based agents, each with a specialized role in creating, managing, and testing traffic scenarios under potential cyberattacks. 

The system defines specific roles and responsibilities for each LLM-based agent to ensure an efficient and comprehensive simulation of connected vehicle traffic models. The Road Network Creation Agent automates the generation of realistic road networks in simulation. The Traffic Scenario Generation Agent creates varied traffic conditions and introduces anomalies like accidents or road closures. The Traffic Control Parameter Agent manages dynamic traffic elements, such as traffic light timings and speed limits. Meanwhile, the Traffic Observation Agent monitors traffic flow, collecting data on congestion and anomalies, which supports the Attack Plan Development Agent in crafting realistic cyberattack strategies. The Attack Implementation Agent executes these strategies within the simulation, ensuring they interact with real-world-like conditions. The Impact Evaluation Agent measures disruptions caused by attacks using metrics like delay and throughput changes, while the Countermeasure Development Agent devises strategies to mitigate identified threats. 

Agents communicate using standardized protocols, ensuring seamless data sharing and coordinated actions. This inter-agent communication creates a feedback loop that enhances both attack planning and countermeasure development. Each agent operates with well-defined input and output parameters, such as the Road Network Creation Agent, which uses map data to generate simulation network files.

A Central Coordination System oversees agent interactions, ensuring synchronized tasks and efficient resource management. This system monitors agent performance and facilitates adaptive role refinement based on simulation needs. The process is inherently iterative, with continuous feedback loops enabling ongoing optimization of cyberattack planning and mitigation strategies. This comprehensive approach ensures robust testing and enhancement of traffic network resilience against evolving cyber threats.

\section{Case Study}
This case study presents an investigation into the application of multi-agent LLMs for creating, attacking, defending, and evaluating a traffic simulation of connected vehicles. The study leverages LLMs to generate Python simulation code, design cyberattack strategies targeting vehicle speed and location broadcasts, develop defense mechanisms, and assess their effectiveness. 
The experiment uses multiple AutoGen agents. Each agent handles a specific task. We tested different APIs, including ChatGPT and Deepseek, to evaluate their performance. This section provides a comprehensive analysis through subsections on data description, experiment design, results, evaluation, and discussion. 

\subsection{Scenario 1: Baseline 5-Vehicle Simulation}

\subsubsection{Data Description}
The dataset includes simulated trajectories for five connected vehicles (labeled 0 to 4) on a 5 km road, with each vehicle’s position, speed, acceleration, broadcasted position, and broadcasted speed recorded at 0.1-second intervals. The simulation parameters, as shown in Table \ref{tab:parameters}, define the physical and behavioral properties of the vehicles and the road. In this simulated scenario, each vehicle broadcasts its speed and location to all vehicles behind it. Each vehicle receives broadcasted speed and location information from all vehicles in front of it to adjust its own acceleration. Also, each vehicle is equipped with sensing capabilities, such as a camera or LiDAR, to measure the speed and location of the immediate vehicle in front of it. Therefore, for the vehicle immediately ahead, it uses the sensed speed and location, while for all other vehicles in front, it relies on the broadcasted speed and location.

\begin{table}[htbp]
\centering
\caption{Simulation Parameters for Connected Vehicle Traffic Model}
\label{tab:parameters}
\begin{tabular}{|p{2cm}|p{3cm}|p{2cm}|}
\hline
\textbf{Parameter} & \textbf{Description} & \textbf{Value} \\
\hline
\(v_0\) & Desired velocity (m/s) & 33.33 (120 km/h) \\
\(T\) & Safe time headway (s) & 1.6 \\
\(a\) & Maximum acceleration (m/s²) & 1.0 \\
\(b\) & Comfortable deceleration (m/s²) & 2.0 \\
\(s_0\) & Minimum distance (m) & 2.0 \\
\(\delta\) & Acceleration exponent & 4.0 \\
\(\text{road\_length}\) & Road length (m) & 5000 \\
 $L$ & Vehicle length (m) & 5 \\
\(n_{\text{vehicles}}\) & Number of vehicles & 5 \\
\(\text{dt}\) & Time step (s) & 0.1 \\
\(\text{entry\_interval}\) & Entry interval (s) & 5 \\
\(\text{hacked\_vehicle\_id}\) & Hacked vehicle ID & 0 \\
$v_f$ & Fake broadcasted speed (m/s) & 0.0 \\
$x_f$ & Fake position offset (m) & -500.0 \\
\hline
\end{tabular}
\end{table}

The IDM calculates each vehicle’s acceleration \(a_{\text{IDM}}\) as:

\begin{equation}  
a_{\text{IDM}}(v, \Delta v, s) = a \left( 1 - \left( \frac{v}{v_0} \right)^\delta - \left( \frac{s^*}{s} \right)^2 \right)
\end{equation}

\noindent where \(v\) is the vehicle’s current speed, \(\Delta v = v - v_{\text{lead}}\) is the relative velocity to the leading vehicle, \(s\) is the actual gap, and \(s^*\) is the desired minimum gap, given by:

\begin{equation}  
s^* = s_0 + v \cdot T + \frac{v \cdot \Delta v}{2 \sqrt{a \cdot b}}
\end{equation}

For connected vehicles, the acceleration is extended to account for multiple vehicles ahead, using broadcasted or sensed data based on adjacency. The connected vehicle IDM calculates the acceleration \(a_{\text{connected}}\) as the minimum acceleration across all leading vehicles, considering their perceived states:

\begin{equation}  
a_{\text{connected}}(v, \Delta v_i, s_i) = \min_i \left[ a_{\text{IDM}}(v, \Delta v_i, s_i) \right]
\end{equation}

\noindent where \(i\) indexes all vehicles ahead of the current vehicle, \(\Delta v_i = v - v_{\text{lead}_i}\) is the relative velocity to the \(i\)-th leading vehicle, and \(s_i\) is the gap to the \(i\)-th leading vehicle, defined as:

\begin{equation}
s_i = 
\left\{
\begin{aligned}
&x_{\text{lead}_i} - x - L && \text{(immediate leader)}, \\
&x_{\text{lead}_i, \text{broadcast}} - x - L && \text{(broadcasted, non-adjacent)}.
\end{aligned}
\right.
\end{equation}

\noindent with $x_{\text{lead}_i, \text{broadcast}}$ being the broadcasted position of the $i$-th leader, potentially modified by the hacker (e.g., $x_{\text{broadcast}} = x + x_f$) and $v_{\text{lead}_i, \text{broadcast}}$ being the broadcasted speed (e.g., $v_{\text{broadcast}} = v_f$). The choice between actual and broadcasted data depends on adjacency: immediate leaders (e.g., Vehicle $i-1$ for Vehicle $i$) use actual $x$ and $v$, while non-adjacent leaders use $x_{\text{broadcast}}$ and $v_{\text{broadcast}}$. 

\subsubsection{Experiment}

The experiment utilizes a multi-agent LLM framework to simulate, attack, defend, and evaluate a connected vehicle traffic system. The framework consists of four specialized LLM agents, each assigned a distinct role. The first agent generates Python simulation code, modeling five vehicles on a 5 km road using the Intelligent Driver Model (IDM). This simulation incorporates vehicle broadcast capabilities and history tracking. The second agent, acting as a hacker, designs a cyberattack strategy targeting Vehicle 0 by manipulating its broadcasted speed to 0 m/s and introducing a position offset of -500 m. The attack aims to exploit the IDM’s gap sensitivity to maximize traffic delay. 

To counter this, the third agent develops a defense mechanism based on decentralized consensus. In this approach, vehicles validate broadcasted data by computing a weighted average of values from neighboring vehicles, with weights determined by proximity. This reduces reliance on potentially compromised data from the hacked vehicle. Finally, the fourth agent evaluates the defense’s effectiveness by comparing traffic delays—measured as the time for all vehicles to traverse the 5 km road—under two scenarios: with and without the defense. Statistical metrics are reported to assess the defense’s performance. 

Table \ref{tab:experiment_design} provides a simplified prompt for each agent, outlining their respective tasks and strategies. Due to size limitations, the table presents a condensed version of the prompts used to generate the simulation code, attack strategy, defense mechanism, and evaluation metrics. 

\begin{table}[htbp]
\centering
\caption{Experimental Design for Multi-Agent LLM Traffic Simulation}
\label{tab:experiment_design}
\begin{tabular}{p{1.7cm}p{3cm}p{3cm}}
\toprule
Component & Simplified Prompt & Description \\
\midrule
LLM Agent 1 (Simulation) & "Create Python IDM traffic simulation for 5 vehicles on 5 km road with broadcast and history tracking." & Generates Python code for IDM-based traffic simulation: \(a_{\text{IDM}} = a \left( 1 - \left( \frac{v}{v_0} \right)^\delta - \left( \frac{s^*}{s} \right)^2 \right)\), with \(s^* = s_0 + v \cdot T + \frac{v \cdot \Delta v}{2 \sqrt{a \cdot b}}\). \\
LLM Agent 2 (Hacker) & "Design cyberattack on Vehicle 0: set broadcast speed to 0 m/s, position offset to -500 m." & Designs cyberattack: sets Vehicle 0’s \(v_{\text{broadcast}} = 0\) m/s, \(x_{\text{broadcast}} = x - 500\) m. \\
LLM Agent 3 (Defender) & "Develop defense: average leader’s broadcast with follower’s sensed values for consensus." & Implements a consensus mechanism: for non-adjacent leaders, calculates \(x_{\text{consensus}} = \frac{x_{\text{lead\_broadcast}} + x_{\text{follower\_actual}}}{2}\) and \(v_{\text{consensus}} = \frac{v_{\text{lead\_broadcast}} + v_{\text{follower\_actual}}}{2}\), averaging the leading vehicle’s broadcasted position and speed with the immediate follower’s actual sensed values. \\
LLM Agent 4 (Evaluator) & "Evaluate traffic delay across baseline, attack, and defense scenarios, report mean time and SD." & Measures traffic delay (time for all \(x \geq 5000\) m), comparing baseline, attack-only, and attack-with-defense. \\
\bottomrule
\end{tabular}
\end{table}

\subsubsection{Experiment Results}

Table \ref{tab:results} presents the mean total travel times for five connected vehicles traversing a 5 km road under three scenarios: Baseline (No Attack), Attack-Only, and Attack + Defense. The Baseline scenario, with a mean travel time of 167.2 seconds. The Attack-Only scenario shows a mean travel time of 184.2 seconds, a 10.2\% increase due to a cyberattack on vehicle 0’s broadcasted data, causing trailing vehicles to perceive a stationary or overlapping leader and triggering emergency braking. The Attack + Defense scenario reduces the delay to 178.1 seconds, a 3.3\% improvement over the Attack-Only scenario, as the defense mechanism averages hacked broadcast data with actual follower data. However, the residual 10.9-second delay compared to the baseline suggests the defense partially restores efficiency, leaving room for further optimization.

\begin{table}[htbp]
\centering
\caption{Traffic Simulation Results Across Scenarios}
\label{tab:results}
\begin{tabular}{p{3cm}p{4cm}}
\toprule
Scenario & Total Travel Time (s, mean) \\
\midrule
Baseline (No Attack) & 167.2 \\
Attack-Only & 184.2  \\
Attack + Defense & 178.1 \\
\bottomrule
\end{tabular}
\end{table}

\begin{figure*}[htbp]
    \centering
    \begin{subfigure}[b]{0.3\textwidth}
        \centering
        \includegraphics[width=\textwidth]{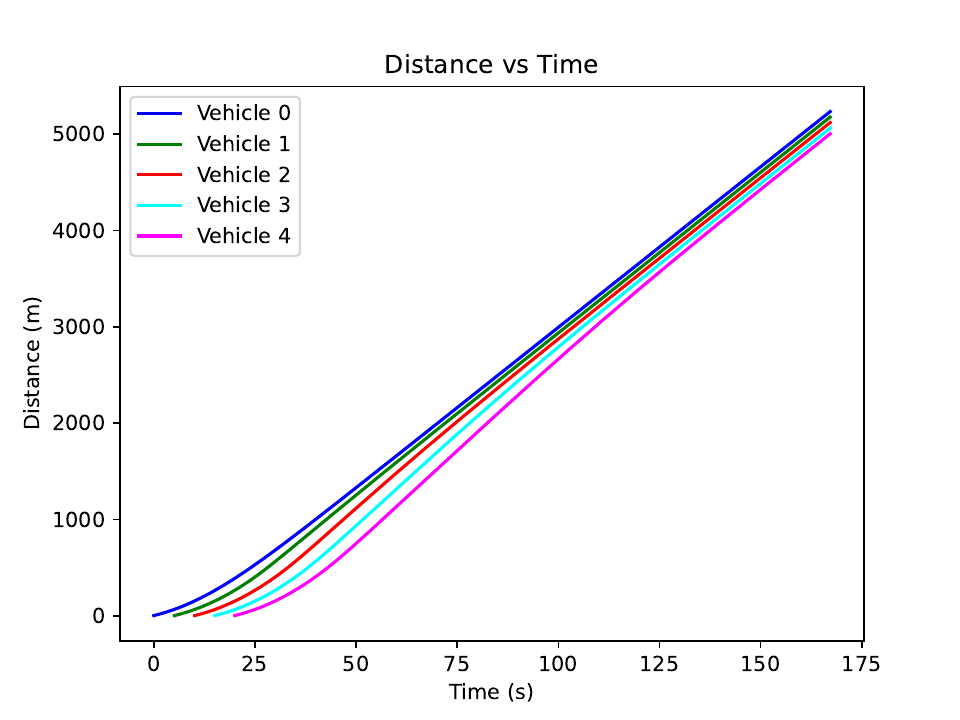}
        \caption{Distance vs Time (Baseline)}
        \label{fig:sub1}
    \end{subfigure}
    \hfill
    \begin{subfigure}[b]{0.3\textwidth}
        \centering
        \includegraphics[width=\textwidth]{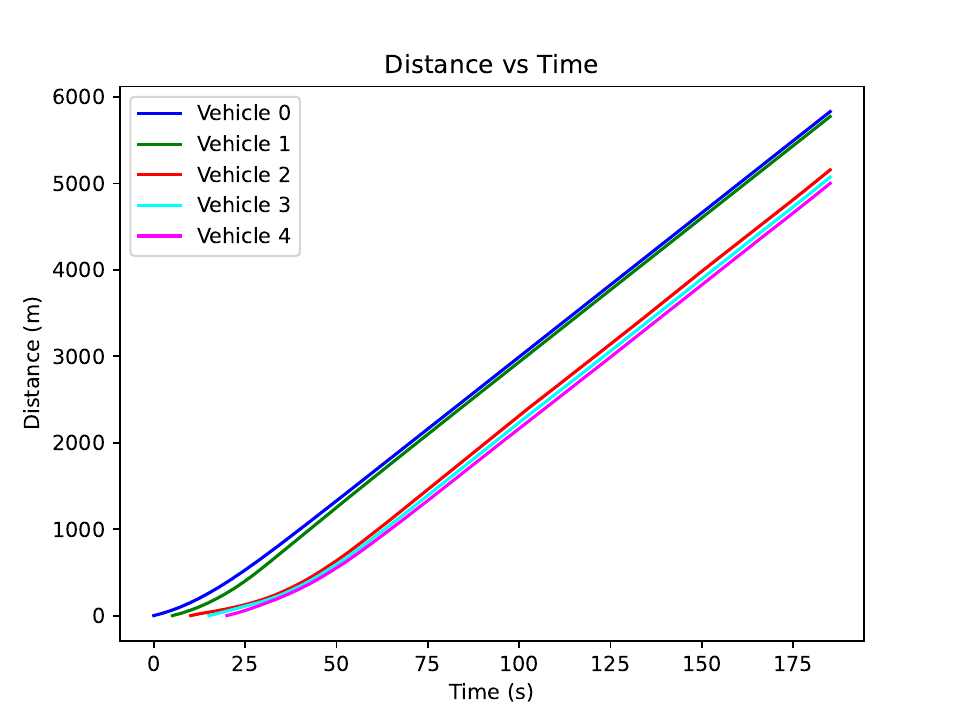}
        \caption{Distance vs Time (Attack-Only)}        

        \label{fig:sub2}
    \end{subfigure}
    \hfill
    \begin{subfigure}[b]{0.3\textwidth}
        \centering
        \includegraphics[width=\textwidth]{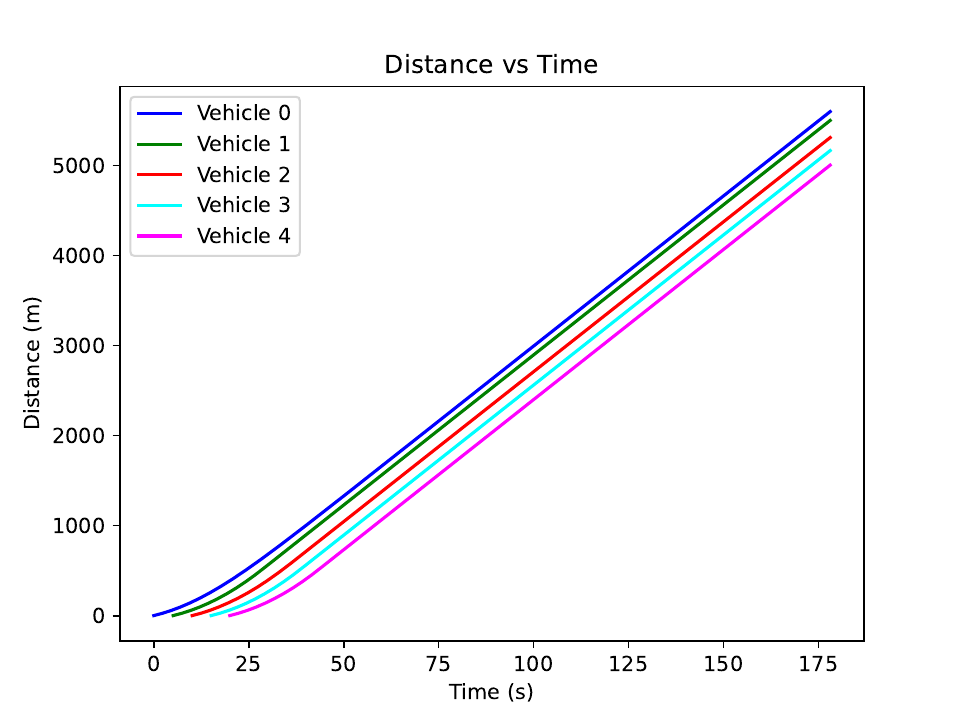}
        \caption{Distance vs Time (Attack + Defense)}        

        \label{fig:sub3}
    \end{subfigure}
    
    \vspace{0.5cm} 

    \begin{subfigure}[b]{0.3\textwidth}
        \centering
        \includegraphics[width=\textwidth]{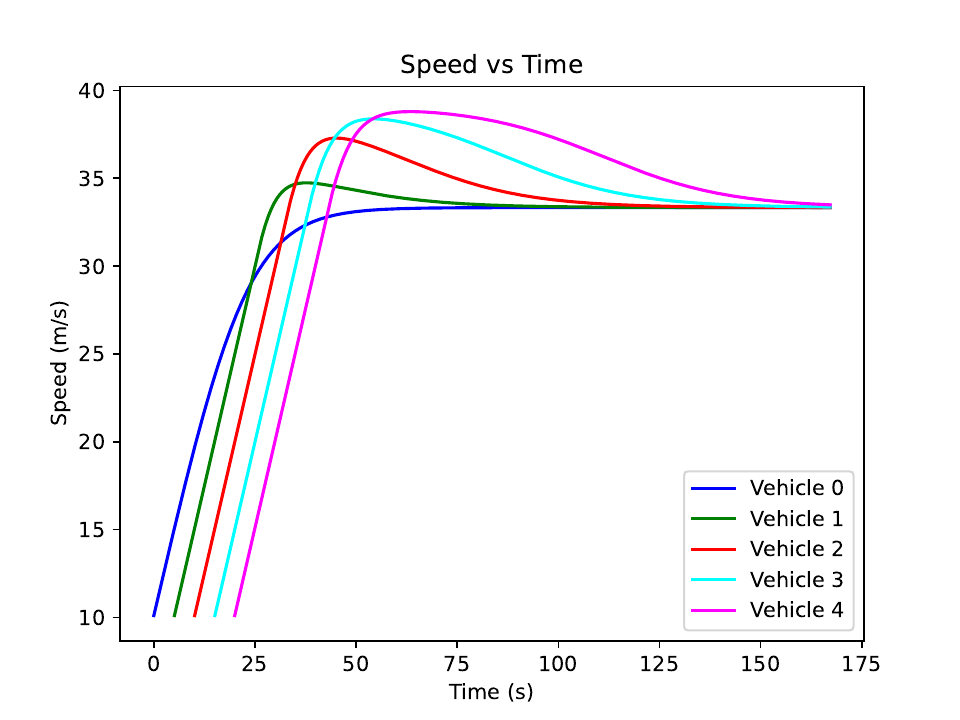}
        \caption{Speed vs Time (Baseline)}        

        \label{fig:sub4}
    \end{subfigure}
    \hfill
    \begin{subfigure}[b]{0.3\textwidth}
        \centering
        \includegraphics[width=\textwidth]{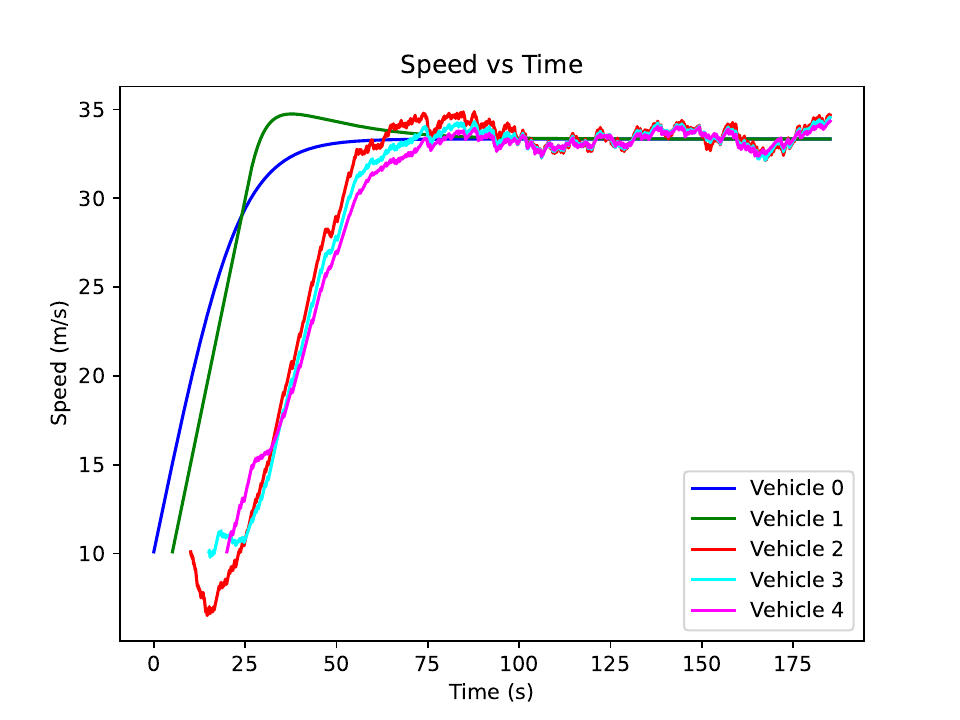}
        \caption{Speed vs Time (Attack-Only)}
        \label{fig:sub5}
    \end{subfigure}
    \hfill
    \begin{subfigure}[b]{0.3\textwidth}
        \centering
        \includegraphics[width=\textwidth]{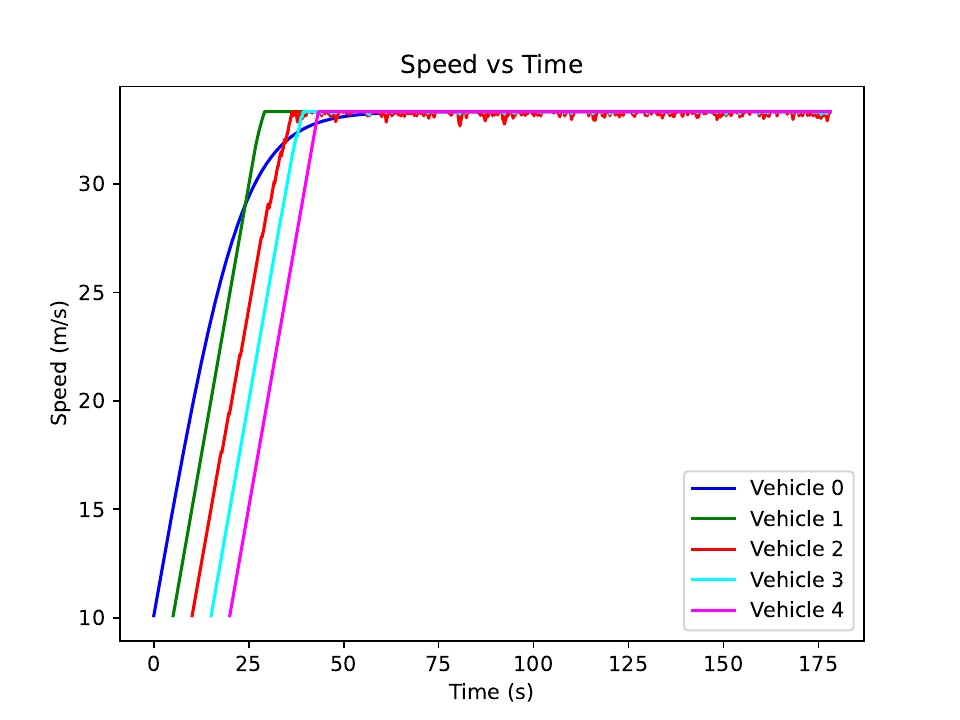}
        \caption{Speed vs Time (Attack + Defense)}
        \label{fig:sub6}
    \end{subfigure}

    \caption{Vehicle Trajectories (Distance vs. Time) for Baseline, Attack-Only, and Attack + Defense Scenarios}
    \label{fig:trajectories}
\end{figure*}

Figure \ref{fig:trajectories} presents a visualization of vehicle trajectories (Distance vs. Time and Speed vs. Time) across three scenarios: Baseline (No Attack), Attack-Only, and Attack + Defense. The figure, organized into six subplots (a–f), illustrates the impact of a cyberattack on the lead vehicle's broadcasted data and the effectiveness of a defense mechanism. Each scenario is represented by a pair of plots: Distance vs. Time (subplots a, b, c) and Speed vs. Time (subplots d, e, f), with vehicles color-coded by ID. The Baseline scenario (subplots a and d) shows smooth, linear trajectories, with all vehicles reaching 5000 meters in approximately 167 seconds. Speed plots confirm steady cruising at ~33 m/s. 

In the Attack-Only scenario (subplots b and e), the cyberattack causes significant disruption. The lead vehicle (Vehicle 0)’s trajectory remains unaffected, but Vehicles 2, 3, and 4 come to a complete stop around 20 seconds due to the manipulated broadcast data. Vehicle 1 experiences a temporary slowdown but recovers. The Speed vs. Time plot reveals that Vehicles 2–4 drop to low speed, while Vehicle 0 maintains its speed. This stoppage, caused by the attack’s manipulation of Vehicle 0’s position and speed, results in a 10.2\% increase in travel time (184 seconds). 

The Attack + Defense Scenario (subplots c and f) demonstrates that the recommended defense mechanism effectively smoothens the speed vs. time curves compared to the attack scenario. Although some fluctuations persist as vehicles reach cruising speed, the overall performance is significantly better than in the attack scenario. This is promising because these results stem from a single prompt; allowing LLM agents to refine this mechanism could potentially lead to an even more robust defense strategy.

\subsection{Scenario 2: Sioux Network}
\label{sec:sioux-case-study}

\begin{figure}[htbp]
    \centering
    \includegraphics[width=0.5\linewidth]{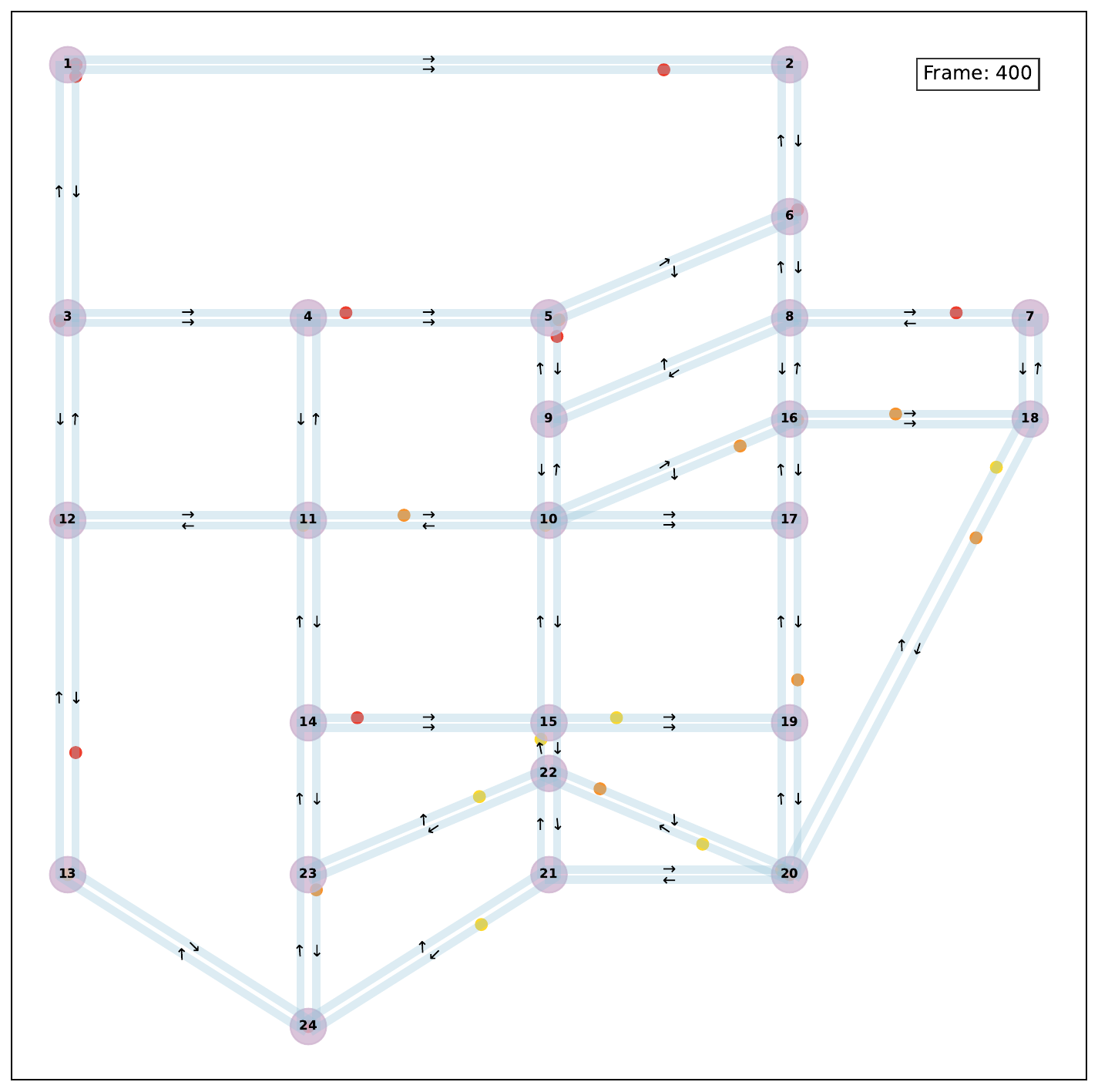}
    \caption{Road layout of the Sioux network with multiple intersections and single-lane links, each governed by all-way stop rules.}
    \label{fig:sioux-network}
\end{figure}

To evaluate scalability beyond a simple 5km road, we applied our multi-agent LLM framework to the well-known Sioux network, which contains multiple intersections and road links. Unlike the single-link case, vehicles in this scenario can dynamically reroute based on real-time broadcast data. Figure~\ref{fig:sioux-network} shows the overall layout of the Sioux network, with single-lane links and all-way stop intersections.

We developed a series of prompts to guide LLM outputs for different components of the traffic simulation framework. For dynamic routing, a prompt was used to enable route decisions based on real-time travel times. The LLM output included Python code for vehicles to broadcast their latest link travel times and update their routes accordingly. Next, a cyberattack prompt instructed the LLM to simulate an attack by replacing vehicles' legitimate broadcasted travel-time data with artificially modified values for selected links, inducing rerouting toward congested corridors. To counter this, a defense prompt applied a consensus-based filtering mechanism that rejects outlier travel times deviating from the majority. The LLM produced a short script to integrate this defense into the simulation loop.

\vspace{0.3em}

The initial runs indicate that the LLM-generated traffic simulation system functions as intended. It successfully models route-choice behavior across a multi-intersection network, simulates a targeted cyberattack by injecting falsified travel-time data, and implements a defense mechanism. The hacked broadcast causes significant travel delays by misleading vehicles onto an ostensibly faster route, but the LLM-generated defense partially mitigates this impact. These results validate the system’s ability to autonomously generate, test, and respond to complex traffic scenarios.


\subsection{Discussions}

The results of this case study shows that the multi-agent llms framework has great potential to generate traffic simulation scenarios, design cyberattacks, and develop corersponding defense mechanism to make the transportation system more robust. The LLM APIs we have tested are able to generate the code quite efficiently. For the first, third and fourth agents who are responsible for generate initial simulation, cyberattack, and evaluation code, the API is able to generate code at the first attempts. For the defense agent, the agent has to went through a few iterations before there is no error message. When agent 2 was asked to choose the best attack on the simulated connected vehicles, the agent is able to choose the first vehicle through reasonsing and pick up the appropriate fake speed and location information to be broadcasted. 

\section{Conclusion}
This research investigates the use of multi-agent LLMs to automate the simulation of traffic scenarios, cyberattacks, and defense mechanisms. The authors developed a multi-agent system utilizing the Autogen and LLM APIs. The results demonstrate that the proposed framework can create autonomous connected vehicle simulation and design both cyberattack and defense strategies. The case study validates the framework's ability to automate these tasks.

However, several limitations and areas for future research have been identified:

    \begin{itemize}
        \item Simulation Complexity: The current simulation scenario is relatively simple. Future work should explore more complex scenarios and extend the simulation to other transportation sectors, such as transit, ports, and aviation.
        \item Simulation Platforms: In this study, the LLM agents directly generated Python code for traffic simulation. Future research could leverage existing traffic simulation platforms like SUMO, with LLM agents generating configuration files for these platforms.
        \item Cyberattack Scenarios: The current study focused on a simple cyberattack that modified broadcasted speed and location data. Future research should explore a wider range of attack scenarios to enhance the robustness of the defense mechanisms.
    \end{itemize}

\section*{Acknowledgment}
The authors would like to thank Sarayu Varma Gottimukkala, a graduate student at the University of Houston, for her contributions in developing the simulation platform in this study.

\bibliographystyle{unsrtnat}
\bibliography{ref}

\end{document}